# Pacloud: Towards a Universal Cloud-based Linux Package Manager

Olivier Bal-Pétré
Technological University Dublin
Dublin, Ireland
olivier.balpetre@gmail.com

Pierre Varlez
Technological University Dublin
Dublin, Ireland
pierre.varlez@gmail.com

Fernando Perez-Tellez
Technological University Dublin
Dublin, Ireland
fernando.pereztellez@tudublin.ie

## ABSTRACT
Package managers are a very important part of Linux distributions but we have noticed two weaknesses in them: They use pre-built packages that are not optimised for specific hardware and often they are too heavy for a specific need, or packages may require plenty of time and resources to be compiled. In this paper, we present a novel Linux package manager which uses cloud computing features to compile and distribute Linux packages without impacting the end user's performance. We also show how Portage, Gentoo's package manager can be optimised for customisation and performance, along with the cloud computing features to compile Linux packages more efficiently. All of this resulting in a new cloud-based Linux package manager that is built for better computing performance.

## 1. INTRODUCTION
Package managers are an essential part of any Linux distribution: they allow to install, update or remove software on any supported device. In this proposal, the Linux software is distributed by using packages containing the executable software as well as metadata files containing various information, including the dependencies to other packages, the version number and other elements useful for the installation or maintenance.

Most Linux distributions use binary packages, that are distributed in pre-compiled packages. Whilst it allows for a fast installation, these packages are compiled for generic hardware and often contain more dependencies than necessary. A good example is the well-known LibreOffice software, that can be used with GTK2/3 or Qt framework. The LibreOffice package is built to be compatible with every user interface framework, hence heavier than necessary: only one framework will be used for this software installation.

To optimise configuration and installation performance, source-based Linux distributions are used, one of the most famous being Gentoo Linux.

Gentoo's package manager (Portage) builds packages from source code and allows for specific compilation flags. This feature allows to have packages that are optimised for a specific hardware. Portage also allows to build and install packages for specific system requirements, with the help of USE flags [4]. These flags are used to specify optional dependencies and settings. Using the previous example, the LibreOffice package can be built with the *gtk* USE flag, thus having dependencies to GTK related packages and not building it to be used with Qt.

Gentoo's package manager is widely regarded as efficient, even though some attempts have been made to improve the interface and package access speed [9]. One of the weaknesses of this package manager is regarding low-spec or portable hardware. It is not designed for this kind of hardware, particularly single-core devices and single-board computers, that have less computing power and rely on lower power consumption. By its nature, compilation is a heavy task, that may require a lot of computing power and it can be slow on weak or single core processors, thus not making Portage and any source-based package manager a good choice for this kind of hardware.

The aim of this research work is to propose an approach to get rid of the compilation load for low-power consumption or single core devices. We propose to use cloud services to pre-build packages on demand and manage them for their distribution, thus giving performance close to binary package managers and the same configurability as source-based package managers. In this proposal, we show how the public cloud, in particular, Amazon Web Services (AWS) allowed us to build a cloud architecture able to compile packages on distributed nodes, store and distribute built packages for users with similar needs. The proposed package manager resulted in speeding up the process of installation and optimising the amount of time and resources required for the compilation tasks.

## 2. RELATED WORK
This research heavily relies on Gentoo's package manager, as the goal is to optimise the compilation phase. To understand how we can improve it, we must first look at how Portage fetches and builds packages.

Portage's repository is a tree structure containing ebuild files [3]. Ebuild files are bash files containing information of the packages,

as well as functions to execute a custom installation. The information includes a description of the package, a URL to the sources, the license, keywords that designate which processor architecture have been tested to work for this package, USE flags and dependencies, both for build-time and run-time. All of this information is used to download, build and install packages with their dependencies.

Portage's dependencies management, including ebuild descriptions and how the database works, is described in a paper by Remco Bloemen and Chintan Amrit [12]. On this paper, they describe a method to parse ebuild files using Paludis and a custom built program to produce the dependency graph for a specific package, and how they used the CVS revision control system where the database is kept to track changes for a specific package. The change over time of the dependency graph is analysed in another paper by the same authors [13].

Portage is a highly configurable package manager, that can be used in different configurations, even on top of existing operating systems. The Gentoo Prefix project [6] allows to install Portage on specific locations and keeps it separate from the rest of the system, thus avoiding conflicts when installed on operating systems including their own package management. In [8], Guilherme Amadio and Benda Xu describe how Gentoo Prefix can be used to install Portage on high performance computing systems, in order to get up to date and well performing packages using cross compiling.

As mentioned in the introduction, some efforts have already been done in improving Portage and making it faster. The given example [9], though, is focused on the fact that emerge, built in Python, and the file-based database are not built for speed but for portability and ease of configuration. The solution they propose, built in C++ and using Berkeley DB, is faster but less portable and less secure.

In our proposed solution, the client side keeps the original principles of portability and ease of configuration which are important aspects in our proposal. In our proposal is also important to speed up the compilation task and reduce the load work on the end-user device. There are multiple existing solution on that regard, some presented in the next paragraphs, but they all displayed some defaults that we try to fulfill.

Portage already allows to have a binary packages build server[5]. However, it requires to have an available server and the USE flags have to be identical to a subset of the USE flags required by the clients. The USE flags limitation makes it unsuitable for multiple clients, as their might be conflicting USE flags. Pacloud, the proposed Linux package manager, gets rid of these limitations: the USE flags are specified by the client and thus we can have binary packages that completely depend on the client's configuration. Pacloud is also built to be used by multiple people, with features that are described in Section 3.1.

There is also a project from a Gentoo developer that aims to create a build service [7]. The Gentoo Build Service project uses a cluster of builder nodes in Docker containers linked to a web interface that manages them. It would have been a great fit for our project, however this project was still on the testing phase. It is still in experimental state as of the writing of this paper. Also, it is built in Go, which does not fit our design choices on portability, as Go requires to be compiled, which would require multiple binaries for each architecture on the client side. Using Python gets rid of this constraint.

Distcc[11], a program designed to distribute compiling tasks across a network to participating hosts, can also be used in Gentoo. It is particularly used when cross-compiling packages. The basic idea is that a client that requires compiling a package will ask servers to perform the task, thus speeding up the compilation. It scales efficiently and works well with Portage. However, it is an elegant solution for a static network configuration, as the workers are called via their IP addresses, but it might not be suitable with cloud solutions involving dynamic IPs. All instances should be launched and we should know their IP addresses in order to be able to add them to the network of participating hosts. It also requires an extra layer of abstraction, as the client is the one ordering the build but we do not want the client to compile anything, given that we have hosts that are designated only to compile.

Some package managers allow users to customise packages. A good example, showing an original way of doing it, is Nix[10], from the operating system based on NixOS. Nix is mostly used to install binaries, but is able to compile packages from Nix expressions as well. A Nix expression is a functional language that describes a package (or *derivation*) and how to build it from source. Each package in the NixOS repositories is build from Nix expressions.

The package manager lets the user modify the Nix expressions to customise the build of a package, as Gentoo Linux does with the ebuilds and USE flags. Then the packages can be pushed to a server to be stored. Other Nix machines can pull the binaries to avoid to compile them too[1]. It is also possible to set a network of Nix machines to compile packages in parallel. The packages will be distributed between each node of the network to be compiled. The binaries will then be shared with the other machines[2]. This approach is close to what Pacloud aims to do as each node could be a user. However, each node needs to be registered on the other machines, making this system hard to use on a large scale and each machine trusts the others to deliver the right package, which raises security issues. Pacloud resolves this problem with a centralised compilation farm on AWS servers. The packages are available for every user, but they are compiled on servers managed by a trusted third party.

## 3. SYSTEM ARCHITECTURE

This proposal is split into two main parts: the server side that was entirely created with the CloudFormation service (YAML format) to be deployed quickly in Amazon Web Services cloud and the client side that was created using Python 3 to be as portable as possible.

## 3.1 Server-Side Architecture

### 3.1.1 Request Process

Pacloud is designed to compile the packages according to the user parameters. The supported parameters are currently the *version* and the *USEflags*. These parameters allow to customise a package. For example, in order to enable the scrolling functionality in the terminal emulator *rxvt-unicode*, the package must be compiled with the *USEflag* "mousewheel". In Pacloud, a binary is unique for every combination of all the parameters cited before, so each time that a parameter differs, a new binary will be compiled.

When a user requests a package through the Pacloud client (#1 on Fig. 1), a request is sent to the API (#2) and forwarded to a Lambda function (#3). This service allows to run some code without provisioning any server. In our case, the Lambda checks in a DynamoDB table (NoSQL database) if the package was compiled previously with exactly the same parameters (#4). If it was, a URL is returned to the client and the package will be downloaded from a S3 bucket and installed on the client side (#5). If the package was never compiled before with the same parameters, a message is sent in a Simple Queue Service (SQS) queue (#6). A SQS queue is a managed service and is highly available, flexible and reliable by design therefore every message will always be processed.

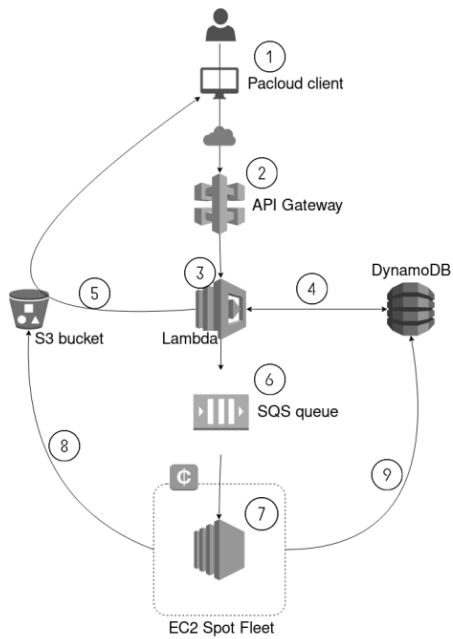

**Figure 1. Server-Side Architecture Design**

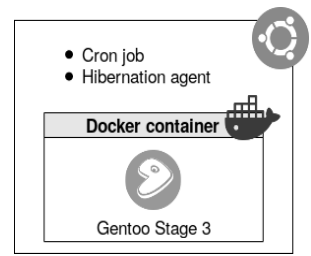

**Figure 2. Spot Instance Detailed**

The computing resources chosen for Pacloud are the Spot instances. They are cheap servers that let you pay only for each minute or second used.

These Spot instances check regularly if there is any compilation request in the queue. If one request is found, it is processed (#7). The queue allows to handle a big amount of requests even if there are not enough Spot instances available at the moment of request. If a message fails to be processed more than twice and the error is not handled in the code of the Spot instance, it will be placed in a other queue. This queue, named dead letter queue, regroups all the messages that cannot be processed to be examined in detail in a later stage.

When a Spot instance compiles a package, the message is invisible in the queue for fifteen seconds. This timer is reinitialised every ten seconds during the compilation. Indeed, setting a few minutes for the timer could be less than the time needed to compile the package, and another instance could begin to compile it too. But a longer timer than the biggest package means that if something fails (i.e. an Availability Zone is down), the message would be hidden for a few hours before being able to be processed by another instance.

Once the package is compiled, the binary is uploaded to a cloud storage, in this case to an S3 bucket (#8) and the meta-data of the binary are stored in a DynamoDB table (#9). Then the message is deleted from the SQS queue. If the compilation failed, the message is also deleted from the SQS queue. However, nothing is uploaded in S3 and instead of the usual meta-data, the error message given by Portage during the compilation is uploaded in the DynamoDB table. This message will be returned to the user that requested the package and any other that tries to compile later the same package with the same parameters.

### 3.1.2 Pacloud Workers

The compilation task usually requires a lot of computing power, which is one reason why this work aims to do it in the cloud. Another reason is the optimisation of power consumption specially in small or low-end devices and powerful compute devices can cost much more. We are also proposing this innovative idea to keep the cost optimised. AWS offers the Spot instances, these instances are a lot cheaper than the on-demand servers and still can be used only for a short and custom duration. Choosing the Spot model can save up to 90% on the on-demand price.

The downside of using Spot instances is that AWS can request back the instances at any time, allowing only two minutes after the notification to finish your job. It may be a short notice to finish some compilations. In order to avoid losing any job in a Spot instance, we use a special functionality on Spot instances called *hibernation*. It allows to hibernate the instance and its associated EBS volume. Once this option is enabled, the instance will hibernate and it will be resumed automatically by the Spot Fleet when an instance will become available again.

The hibernation option has some restrictions[14]. It is available only for the Spot model and on the types of instance C3, C4, C5, M4, M5, R3, and R4. Furthermore, it needs to launch on every instance a background agent named "hibagent" provided by AWS. This agent handles the AWS Spot Fleet notifications and hibernates the machine in a way to be able to resume it later. Currently, this agent is available only for the Linux distributions Ubuntu and Amazon Linux.

The chosen OS in this proposal is Ubuntu Linux (Fig. 2). The AMI contains only the hibernation agent and a Cron job that regularly launch a Docker container. The container is a Gentoo stage 3 container to get Portage. This container image is the minimal installation of Gentoo to get something operational. When run, the container checks in the SQS queue if there is a request, and processes it if found. If nothing was found, the container is removed. The container is also removed after each compilation. This allows to always get exactly the same environment and also to ensure that each package is compiled in the same conditions and will not be affected by the previous compilation.

### 3.1.3 Package Compilation

As we use the Gentoo repositories to get a package database with the ebuilds, we also use Portage to compile the packages. The compilation command is described in Fig 3. It is made up of two

```
env USE="$useflags" \
    emerge --onlydeps --onlydeps-with-rdeps n \
    =$package-$version \
    && emerge --buildpkgonly =$package-$version
```

**Figure 3. Compilation Command**

parts: the compilation and installation of the build-time dependencies on the server, and the creation of the binary for the requested package. The build-time dependencies, opposed to the run-time dependencies, are the packages required to be able to compile the package.

The *env USE="$useflags"* sets the USEflags for current command. The parameter *--onlydeps* allows to compile just the dependencies and not the package of the command and *--onlydeps-with-rdeps n* specifies that the run time dependencies do not need to be compiled. Indeed, the run time dependencies are resolved on the client side, and the client asks the server each dependency, so the server does not need to manage that. Finally, *=$package-$version* specifies the package and the version wanted. The package variable here represents the category and the package name. It is at the format *category/name* like in the Gentoo repositories. The category being the family of packages (e.g. *app-editors*) and the name being the package's name (e.g. *vim*).

The last line *&& emerge --buildpkgonly =$package-$version* builds a binary package only for the package and does not install the package. Indeed, the installation is not needed in our case. The use of *&&* instead of another command makes sure that if the first command fails, this one will not be executed.

### 3.1.4 Computing Resources Architecture

Except for the EC2 instances, all the other AWS services used for the server architecture are managed services supported by AWS. Although we used AWS, this proposal is not limited to use only this cloud provider. This project can also be implemented in other public clouds as we used services supported by other cloud providers. The managed services used are highly available, scalable and reliable by design.

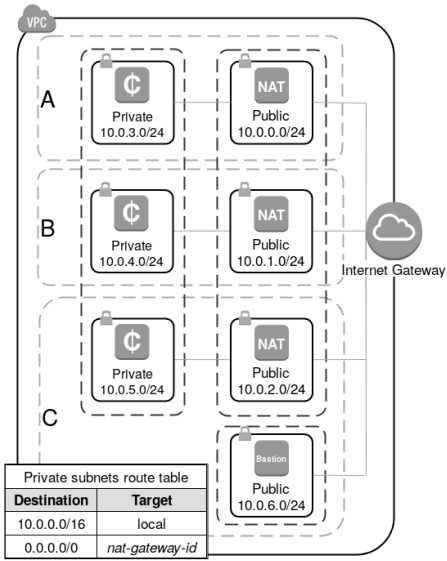

**Figure 4. Computing Resources Architecture**

The most cost-effective solution for the computing services were the EC2 Spot instances category, but they are not highly available, scalable and reliable by default. However, it is possible to get these behaviours with an ad-hoc architecture.

As shown in Fig. 4 the Spot instances are in three Availability Zones (AZ). The Spot Fleet will launch the instances in the three AZs in order to get high availability. This means that if one AZ is down, all the compilation requests interrupted will be available again in the SQS queue and will be processed by instances in another AZ. The NAT instances are also EC2 based and managed by the user, so there is one in each AZ.

NAT instances were chosen instead of the AWS managed NAT Gateway to reduce the cost. They are not implemented with redundancy in an AZ, but enabling EC2 auto recovery on the NAT instances will repair them if they do not pass the health check and will not leave the Spot instances without an Internet connection for more than a few minutes. Their bandwidth is also lower but it is enough for our needs.

**Figure 5. Inbound Rules for NAT Instances Security Group**

| Protocol type | Port number | Source IP |
|---|---|---|
| TCP | 80 | spot-instances-sg |
| TCP | 443 | spot-instances-sg |

**Figure 6. Inbound Rules for Spot Instances Security Group**

| Protocol type | Port number | Source IP |
|---|---|---|
| TCP | 22 | bastion-sg |

To improve the security of the Spot instances, they are in private subnets. However, they still need an access to the Internet to download the source code to compile the packages. NATs were added for this purpose. They allow connections to be initialised from the inside but not from the outside (Fig. 5). To allow us to access them by SSH to perform maintenance, a bastion server was added. The bastion is in a public subnet and allows SSH connections from a particular IP address. The Spot instances allow SSH only from the bastion (Fig. 6).

The AWS managed services used in this proposal are API Gateway, S3, Lambda, SQS and DynamoDB which scale automatically. However, we need to scale the computing resources as well. Scaling the resources out and in according to the number of packages to be compiled in the SQS queue could be a possibility. It could be done easily implementing auto-scaling feature, but this does not answer well the current necessity.

Indeed, as some packages need a few seconds to be compiled, some others require a longer time. This is the first reason why it is difficult to create auto-scaling rules now. While the number of users is small, scale according to this number could result in too many servers started if even just one user requests a lot of quickly compiled packages. Likewise, not enough servers will be started if only slowly compiled packages are requested. The second difficulty is that auto-scaling rules are based on parameters as the time that a user can accept to wait for a compilation or the price that we can pay for the servers. These parameters will guide the auto-scaling choices, but they can be set only according to an economic model to put the application in production. Because of these reasons, the auto-scaling of the computing resources is not part of our proposal at this point.

### 3.1.5 Computing Resources Cost

The cost of the cloud architecture is difficult to evaluate as it will depend of the number of packages to store and of the number of compilation requests. In order to give an idea of the cost, we will split the problem in two: the cost of storage for a given example and the cost of the computing resources for minimal service. All the prices are relative to the date of writing of this paper. All the services are "on-demand" services and are paid only for what we use.

The storage includes the S3 service to store the packages and the DynamoDB service to index the packages stored there. The cost of DynamoDB will be neglected here as it would cost less than 5$ a month for a million read request, a million write request and 10 GB of storage. These numbers are very high for the use we do of it and if they are reached once, the cost will be extremely low compared to the cost of the other services involved. As S3 is an on-demand service, the price will also change depending on the number of packages to store. Based on 360 packages, we estimate the size of a compressed package to 2 MB. Based on this statistic, we can calculate that storing one version of each of the 20,000 packages of the Gentoo's repositories would cost around 40$ a month. Of course, this number can be higher as each package has multiple versions and multiple compilation configurations due to its *USEflag*.

The estimation of the computing resources will neglect the cost of three services: API Gateway, Lambda, and SQS. Indeed, these services cost respectively 3.50$ a million request, 0.20$ a million request and 0.40$ a million request. The remaining service is the EC2 spot instances. As the number of instances would change a lot depending on the needs and the number of users, we will only provide the current price for an EC2 spot instance *c5.2xlarge*. This instance has 8 vCPUs, 16 GiB of memory and costs around 98$ a month if up 24 hours a day.

As we can see, most of the cost resides in the computing resources. A viable economic model is not part of this proposal but we guess that it would need many users to be able to use most of the time of the EC2 instances and be worth.

## 3.2 Client-Side Architecture

The client side is written in Python 3, using as few dependencies as possible. The goal here was to achieve maximum portability and have a single library that works on all kind of Linux distributions as well as different architectures. To that end, it uses a multi-layer architecture: interfaces are built on top of the library that is composed of an API visible to the interfaces and a lower level architecture containing basic elements that are put together to form the API, as described in Fig. 7.

### 3.2.1 Interfaces

Pacloud currently uses a command line interface. As described in Fig. 7, the interface is built on top of the API layer, making it easy to create new interfaces (such as a web interface). It provides a mean for the end user to interact with the basic features of the package manager: searching, installing, removing and upgrading packages, as well as updating the local database.

On the command line interface, arguments are made to be POSIX and GNU compliant: they can be called either with a short version (POSIX compliant) or a long version (GNU compliant). For instance, to search for a package, the user can run *pacloud -s foo* or *pacloud --search foo*, as we can see on Fig. 8. The arguments keywords were made to be readable and describe the functions they are calling, unlike some package managers (mostly Arch Linux's package manager, pacman) that have abstract arguments names.

### 3.2.2 Configuration

The configuration file is by default situated in */etc/pacloud/pacould.conf*. It contains multiple variables, separated in categories in an INI format. The existing categories are local, server and user.

The local category contains the path to the local database (by default, */var/lib/pacloud/db/*) as well as the log files (by default, */var/lib/pacloud/pacloud.log*).

The server category contains a variable for the API Gateway URL and another one for the S3 bucket where the packages are located.

The user category contains USE flags, as well as variables for architecture and compilation flags.

The configuration is read in a Python script, but it is never modified by the program.

### 3.2.3 Local Database

The local database is a flat file database. It uses plain text files in order to make it easy to read and modify for the user. As Portage is used in the server, the local database uses the Gentoo repository tree hierarchy architecture. Every category directory contains packages directories, that contain metadata files.

These metadata files have been translated from Gentoo's ebuilds to JSON files, containing the information we need in a more readable format for our purposes. Every file contains the name, description as well as every version and its associated runtime dependencies. When a package is installed or some packages depending on it are installed, fields are added to this file by the package manager's low level layer. Adding a variable with packages that depend on this package allow us to get a clean list of packages to remove in order to leave a clean installation when we want to remove a package with its dependencies, in case the dependencies are not used for some other package.

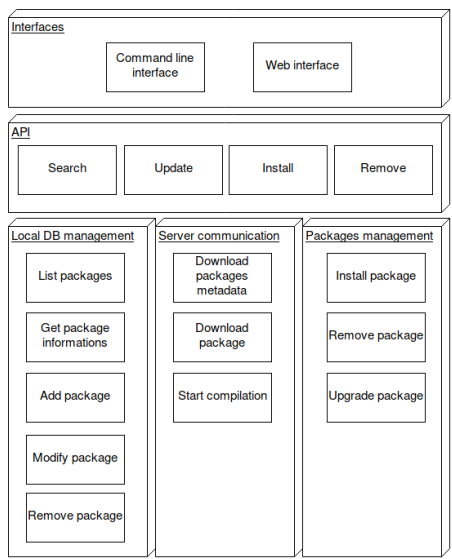

**Figure 7. Client-Side Architecture**

**Figure 8. Searching for a package with the command line interface**

When a package is installed, its built archive is downloaded and put into the directory corresponding to its package, in the local database. The cached version is then re-used if we want to re-install it, thus saving the server some processing and avoiding to download the package multiple time.

### 3.2.4 Dependencies Management

As mentioned earlier, we used Gentoo's repositories and specifically the ebuilds to get a working configuration and compile the packages. The ebuilds have been parsed and translated into JSON files containing information we use to resolve dependencies and have other information on packages. The JSON format is easier to read and process, as it is accepted by virtually any language and framework, while the ebuilds are hard to parse: their format is readable by bash, but because of the use of variables, reading them with Python is a bit challenging.

The dependencies are listed on the ebuilds in different variables: *DEPEND* gives the packages needed to compile and install the package while *RDEPEND* gives the packages needed at run time. The latter is the one we want to use on the client side, as the other one is mainly used to be able to compile the packages, hence on the server side. This is why on our JSON formatted metadata files, we get the value of *RDEPEND* in our *dependencies* field.

A first parsing is performed when the JSON file is created, giving us a list of strings for the dependencies. If there is no dependency using a USE flag, it eases up the process, as we only need to process the dependency string on itself.

On the client side, two main operations are performed when checking for dependencies: checking the USE conditional dependencies and finding the right version.

The USE conditional dependencies work this way: if a dependency needs to be installed because a USE flag is set, the syntax will be the following: *use? ( dependency )* where *use* is the USE flag and *dependency* is the dependency string. What makes it harder to parse is that this kind of conditions can be nested, which requires a recursive algorithm to process every level of the dependency.

The dependency strings are following the format *SpecifierPackage-Version*, where the specifier is a relationship operator (>=, >, ~, =, <=, <), package is the name of the package including its category and the version is the version of the dependency. Parsing it allows us to perform a check on the local database to see if a version fits the criteria. If it is the case, it will be the version we use in the dependency management.

### 3.2.5 Server Interaction

As it was previously noted, the server address is written in the configuration file. The communication with the server uses HTTP requests. For that, the *urllib.request* Python module is used.

The package manager downloads different files from the server side:

When the local database is being updated, a *manifest.txt* file, situated at the root of the S3 bucket, is downloaded. This file contains a list of all categories in the database.

Every category in the manifest file are located on the S3 bucket and downloaded in JSON format.

The API gateway is called when a package has to be installed. The request for the package contains the package name, version and some configuration parameters (mainly, the USE flags). The server then responds with a JSON formatted response, that is parsed and, depending on the value of the "status" field, the package is downloaded from the URL included in the response or we will wait for the package to finish compiling.

The waiting phase is currently a simple sleep, after which a new request is made. An improvement for that would be to open a connection with the compiling server, for instance with a socket, and wait for a response saying that the package has been compiled. This method would also allow to get more information from the compilation.

The main issue is that the compilation servers are in private subnets and access the Internet through NAT instances for security reasons. It means that the client cannot create the socket with the server. Initialise the socket from the server is not a good idea neither as the compilation doesn't always begin just after the request, so the client could not be reachable on the same IP for example. The best solution could be to initialise a socket from the client and another one from the compilation server to a publicly exposed server. This server would have the role to link the sockets of the client and server to allow the server to send a live status to the client requesting the compiling package. However, this feature is not in the scope of this work.

## 4. SYSTEM PERFORMANCE

To calculate system performance, we have analysed the installation time of a heavy package (*gcc*) as well as a lighter package (*ncurses*). We will compare the installation time on Pacloud running different types of instances, as well as Portage running on a Raspberry Pi 2 and different laptops: a high end laptop (Dell XPS 15 9560, using an i7-7700HQ @ 3.8 GHz), a mid-range laptop (Asus Swift 3, using an i5-7200U @ 3.1 GHz) and an old gaming laptop (Asus G53S, using an i7-2630QM @ 2.9 GHz). On the client side, the Pacloud tests are run on a t2.micro EC2 instance using Ubuntu.

The difference between compiled and binary packages performances has already been studied [15], for that reason we will not perform any tests on that matter and focus on the performance of compilation and installation, comparing Pacloud to Portage.

Performance using Pacloud, as seen on the tables on Fig. 9 and 10 and the chart on Fig. 11, are calculated on a Docker container in a t2.micro EC2 instance with Pacloud installed on Ubuntu, thus allowing for the same hardware and software for all these tests. We should also note that the cross-architecture compiling has been shown to be doable with a proof of concept but it has not been implemented in Pacloud at the time of the writing of this paper. As such, the results regarding the Raspberry Pi 2 for Pacloud should be a bit slower, because of the cross-compilation (about 10% slower).

Our tests show that for a large package such as *gcc*, a high-end laptop will be very competitive on the compilation time: the largest kind of EC2 instances we tested (c5.9xlarge, with 36 cores at 3.0 GHz) is only 13% more efficient than a Dell XPS 15 9560 (with an i7-7700HQ, with 8 cores at 4.2 GHz), as we can see on Fig. 9. The high-end laptop is more efficient running portage than Pacloud with most EC2 servers.

**Figure 9. Benchmark gcc-6.4.0-r1**

| Cores | Clock (GHz) | Machine | Time |
|---|---|---|---|
| 4 | 0.90 | Raspberry Pi 2 | 08:12:51.00 |
| 2 | 3.50 | c3.large | 01:51:19.42 |
| 2 | 3.00 | c5.large | 01:18:23.52 |
| 8 | 2.90 | Asus G53S | 56:13.26 |
| 4 | 3.10 | Acer Swift 3 | 52:29.78 |
| 4 | 3.00 | c5.xlarge | 46:10.89 |
| 8 | 3.00 | c5.2xlarge | 33:30.77 |
| 8 | 3.80 | Dell XPS 15 9560 | 28:35.84 |
| 16 | 3.00 | c5.4xlarge | 25:35.89 |
| 36 | 3.00 | c5.9xlarge | 24:53.34 |

**Figure 10. Benchmark ncurses-6.1-r2**

| Cores | Clock (GHz) | Machine | Time |
|---|---|---|---|
| 4 | 0.90 | Raspberry Pi 2 | 26:00.92 |
| 2 | 3.50 | c3.large | 4:26.85 |
| 2 | 3.00 | c5.large | 3:01.35 |
| 4 | 3.00 | c5.xlarge | 2:49.18 |
| 8 | 3.00 | c5.2xlarge | 2:48.92 |
| 16 | 3.00 | c5.4xlarge | 2:31.17 |
| 8 | 2.90 | Asus G53S | 2:05.59 |
| 36 | 3.00 | c5.9xlarge | 2:02.86 |
| 4 | 3.10 | Acer Swift 3 | 1:53.38 |
| 8 | 3.80 | Dell XPS 15 9560 | 1:38.52 |

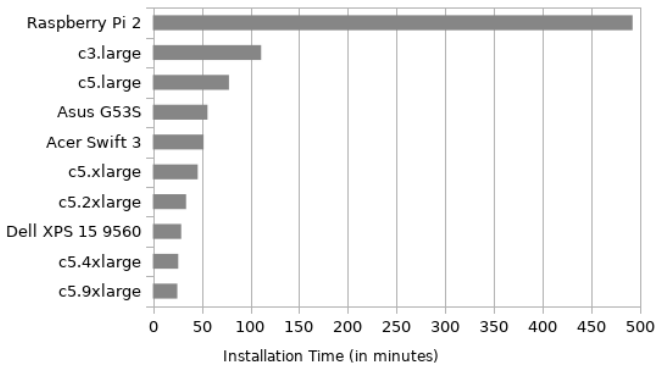

**Figure 11. bar Plot Benchmark gcc-6.4.0-r1**

The mid-range laptop (Acer Swift 3, 4 cores at 3.1 GHz) and the old laptop (Asus G53S, 8 cores at 2.90 GHz) have similar performances and most instances are more efficient. As such, we can halve the time needed to install *gcc* on these laptops by using a c5.9xlarge or a c5.4xlarge instances.

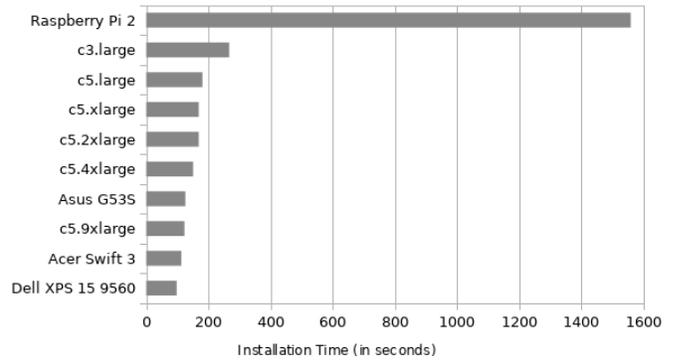

**Figure 12. Bar Plot Benchmark ncurses-6.1-r2**

The most interesting part is with the Raspberry Pi 2: on its own, it took more than 8 hours to compile gcc. Using Pacloud with the largest servers we have would allow to install it in only 5% of that time.

Regarding *ncurses*, except for the largest tested EC2 instance, laptops have better performances, as we can see on the table on Fig. 10 and on the chart on Fig. 12. That can be explained by multiple elements:

- Pacloud adds operations to what Portage does: we need to send a query to the compilation server, then download the compiled archive, which takes time.
- The client side of Pacloud is not entirely optimised yet, especially regarding the time between the end of the compilation and the download of the package.
- Having large servers is not very useful for small packages: there is less compilation involved so we benefit less from having a large number of cores, but having a higher clock rate is important, which is why the high-end laptop performs so well here.

The Raspberry Pi still performs poorly compared to Pacloud, allowing us to reduce down to 7.9% of the current time required to install *ncurses*.

Although performances for a high-end laptop are equivalent whether we use Gentoo's package manager or Pacloud, and laptop performances are usually better for small packages, Pacloud has multiple pros that these benchmarks do not reflect:

- The load during the compilation is not exerted on the laptop but on the cloud instances, which means we save a lot of battery for the laptop as well as using the local hardware less (which means a longer lifespan for the laptop or the end device) and being able to use the laptop normally.
- We can compile multiple packages in parallel by scaling our instances out, which means our total compilation time is highly reduced when we need to compile multiple packages.
- Once a package is successfully compiled for the user configuration, it will not need to be compiled again, which means that the more people use Pacloud, the shortest the installation time would be.

In order to show a more useful and real-life example, we made a benchmark on a number of well-known packages as shown on negligible overhead. We finally get the 16 packages installed at the same time that the longer package to install. Here we get a time similar to the installation time of *gcc-7.3.0-r3* for the installation of all of the packages. It is important to note that with a higher number of packages than servers, we could get the same installation time. Indeed, the compilation of the 17[th] package should begin after the compilation of the first package compiled. If the sum of its compilation time and the compilation time of the 17[th] is less than the time of the longest package, the total time will still be the same.

Fig. 13. The benchmark was made one package at the time, then with the 16 packages at once with 16 servers up, forcing a parallel compilation. As expected, we can see that the installation time for a unique package as *gcc-6.4.0-r1* still is similar than before. Indeed, the compilation is made in parallel on another server with

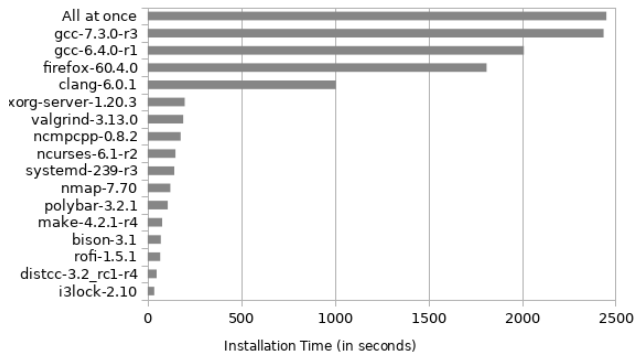

**Figure 13. Bar Plot Benchmark – Packages Compiled on EC2 c5.2xlarge**

## 5. CONCLUSION AND FURTHER WORK

Using a cloud architecture allows us to provide users with an innovative package manager that provides both compute optimisation and easy installation of packages in Linux distributions. All this was possible by having the heavy compilation process done in the cloud. The biggest advantage of using the cloud instead of a simple server is that it is more reliable and can support a lot of users by scaling depending on the load that the servers are subject to.

Although we have a working solution in the early stages, it is important to note that this package manager is being improved daily. There are elements still needed to be worked on, such as the support of more architectures through cross-compilation, which has been proved to be doable but not yet implemented, or the support of compiler flags to provide more optimised packages depending on the user specifications.

Most importantly, this research work showed us that the cloud can improve package managers as they are seen today, to provide packages that rely on the users' needs and not on what maintainers think are the best for the most people. It is all about giving to Linux users the choice to select the customisation that they want, which is in pair with the philosophy of free software.

## 6. REFERENCES

[1] Eelco Dolstra, Merijn de Jonge, Eelco Visser. 2004. Nix: A Safe and Policy-Free System for Software Deployment. *18th Large Installation System Administration Conference (LISA 04)*(Atlanta, Georgia, USA. USENIX, November 2004.)*, 79-92*

[2] Eelco Dolstra. The Purely Functional Software Deployment Model. *PhD thesis* (January 2006) Faculty of Science, Utrecht, The Netherlands. ISBN 90-393-4130-3. 115-118

[3] Gentoo Foundation. Gentoo Linux Development Guide. Retrieved May 30, 2019. https://devmanual.gentoo.org/ebuild-writing/file-format/index.html

[4] Gentoo Foundation. 2012. *Gentoo wiki.* (2012) Retrieved May 30, 2019 https://wiki.gentoo.org/wiki/USE_flag

[5] Gentoo Foundation. 2014. Binary Package Guide. (2014). Retrieved May 30, 2019 https://wiki.gentoo.org/wiki/Binary_package_guide#Setting_up_a_binary_package_host

[6] Gentoo Foundation. 2015. Gentoo Prefix. (2015). Retrieved May 30, 2019 https://wiki.gentoo.org/wiki/Project:Prefix

[7] Gentoo Foundation. 2018. Gentoo Build Service. (2018). Retrieved May 30, 2019 https://wiki.gentoo.org/wiki/Project:Build_Service

[8] Guilherm Amadio, Benda Xu. 2016. Portage: Bringing Hackers' Wisdom to Science (2016)

[9] Simon Maynard. 2005. Package Management System for Gentoo Linux. (2005)

[10] Nix. 2018. Nix. (2018). Retrieved May 30, 2019 https://nixos.org/nix/

[11] Martin Pool. 2004. distcc, a fast free distributed compiler. (2004)

[12] Remco Bloemen, Chintan Amrit, Stefan Kuhlmann, Gonzalo Ordóñez-Matamoros. 2014. Gentoo Portage Package dependencies (2014)

[13] Remco Bloemen, Chintan Amrit, Stefan Kuhlmann, Gonzalo Ordóñez-Matamoros. 2014. Diffusion in Open Source Software. (2014)

[14] Amazon Web Services. 2017. Amazon Elastic Compute Cloud User Guide. Retrieved May 30, 2019 https://docs.aws.amazon.com/AWSEC2/latest/UserGuide/spot-interruptions.html

[15] Christopher Smart. 2009. Gentoo Optimizations Benchmarked. Retrieved May 30, 2019 http://www.linux-mag.com/id/7574/